\documentclass[12pt,preprint]{aastex}

\usepackage{natbib}

\shorttitle{Evidence of tidal debris from $\omega$ Cen in the Kapteyn Group}
\shortauthors{Wylie-de Boer et al.}

\begin{document}

\title{Evidence of tidal debris from $\omega$ Cen in the Kapteyn Group}

\author{Elizabeth Wylie-de Boer\\ Ken Freeman}
\affil{Research School of Astronomy and Astrophysics, Australian National University, \\ Mount Stromlo Observatory, Cotter Rd, Weston Creek, ACT 2611, Australia}
    \email{ewylie@mso.anu.edu.au; kcf@mso.anu.edu.au}
\author{Mary Williams}
\affil{Astrophysikalisches Institut Potsdam, \\An der Sternwarte 16, D-14482 Potsdam, Germany}
    \email{mary@aip.de}
    
\begin{abstract}
This paper presents a detailed kinematic and chemical analysis of
16 members of the Kapteyn moving group.  The group does not appear
to be chemically homogenous.  However, the kinematics and the chemical
abundance patterns seen in 14 of the stars in this group are similar
to those observed in the well-studied cluster, $\omega$ Centauri.
Some members of this moving group may be remnants of the tidal
debris of $\omega$ Cen, left in the Galactic disk during the merger
event which deposited $\omega$ Cen into the Milky Way.

\end{abstract}


\section{Introduction}

The idea of hierarchical galaxy formation, in which galaxies are
believed to form from the aggregation of smaller elements (see
review by \citet{Freeman02}), has been around since \citet{Searle78}
first proposed this theory as a challenge to the belief that galaxies
formed through the smooth collapse of a large protocloud \citep{Eggen62}.
The identification of debris from these smaller fragments remains
of utmost importance in modern studies of theoretical and observational
stellar dynamics.

The most massive Galactic globular cluster, $\omega$ Cen, has several
unique physical properties which suggest that there are very
significant differences in star formation histories, enrichment
processes and structure formation between $\omega$ Cen and other
normal globular clusters \citep{Bekki03}.

A commonly accepted scenario of formation of $\omega$ Cen is that
it is the surviving nucleus of an ancient dwarf galaxy, the outer
envelope of which was entirely removed by tidal stripping as it was
accreted by the Galaxy (\citet{Bekki03} and references therein).
Through numerical simulations, \citet{Bekki03} demonstrated the
dynamical feasibility of $\omega$ Cen forming from an ancient
nucleated dwarf galaxy which was accreted into the young Galactic
disk.  

\begin{figure}
\plotone{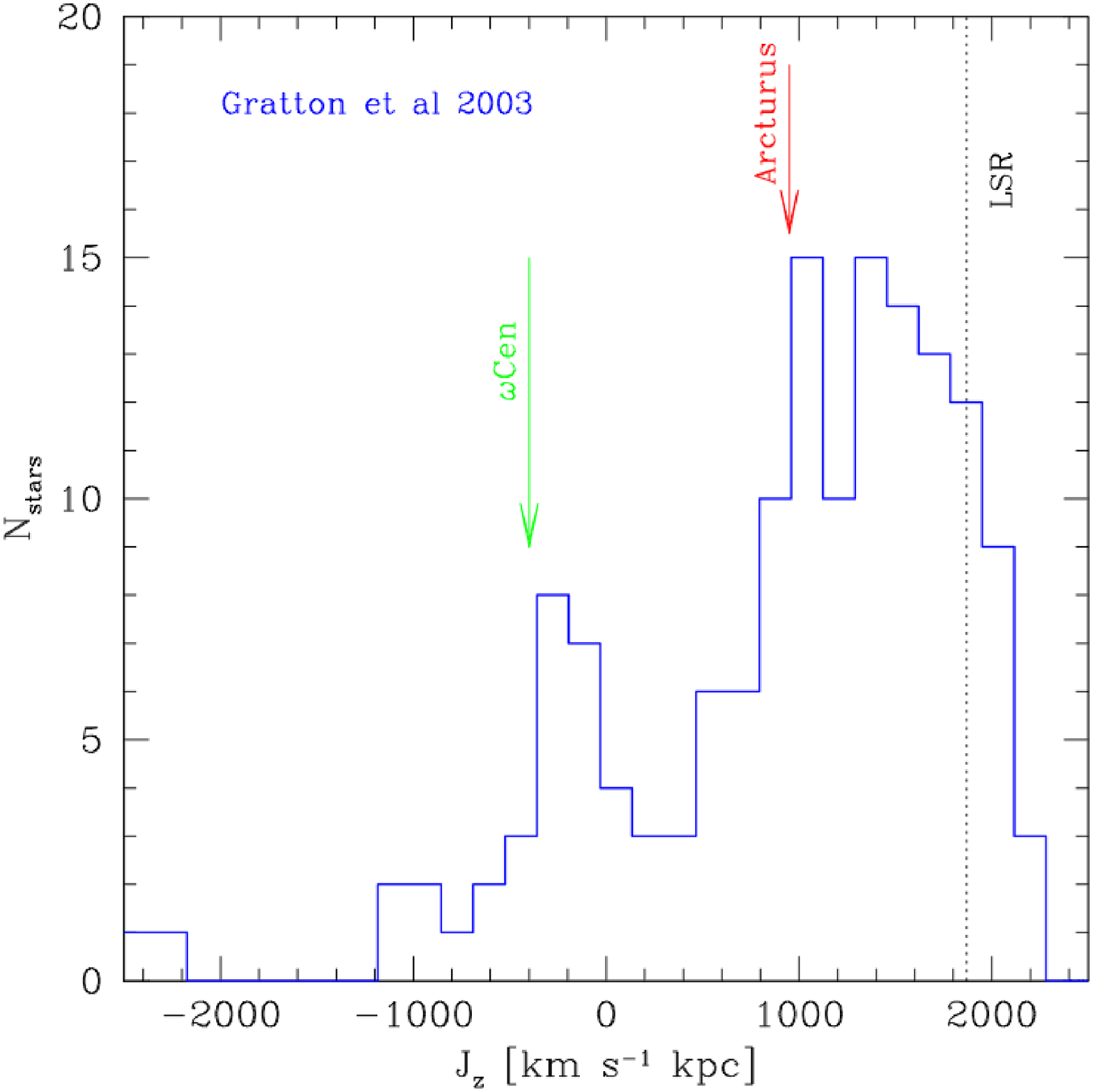}
\caption{Distribution of specific angular momentum of all stars in
the metal-poor compilation of Gratton et al. (2003).  The excess
of stars corresponding to the $\omega$ Cen group is evident. 
(Taken from Meza et al. 2005.)\label{mezafig}}
\end{figure}

\citet{meza05} used numerical simulations to investigate the
characteristics of tidal debris from satellite galaxies. They showed
that these satellites deposit a large fraction of their stars into
either the disc component of the Milky Way or into the halo, showing
distinct ``trails" in the angular momentum -- energy plane, depending
on the plane of the satellite's orbit during disruption.  Meza et
al.  discussed the presence of the $\omega$ Cen stellar moving group
(i.e. the stellar debris) in two studies of metal-poor stars in the
solar neighbourhood, those of \citet{Beers00} and \citet{Gratton03}.
Figure \ref{mezafig} (taken from \citet{meza05}) shows how the
$\omega$ Cen group is distinguished in the angular momentum
distribution for the Gratton et al. (2003) sample.  In both this
sample and the Beers et al. (2000) sample, the $\omega$ Cen group
appears as an over-density of stars at very specific rotational
velocity or angular momentum values.  The radial (U) velocities for
the $\omega$ Cen candidate stars are observed to have a symmetric
distribution, ranging from -300 to 300 km s$^{-1}$.  An earlier
simulation by \citet{Bekki03} of the accretion
of the $\omega$ Cen parent galaxy showed a strong plume of debris
stars in the solar neighborhood with L$_z$ near $-500$ kpc km
s$^{-1}$.  See also the discussion by \citet{Mizutani03} on the kinematics of tidal debris from the parent galaxy.

\begin{figure}
\plotone{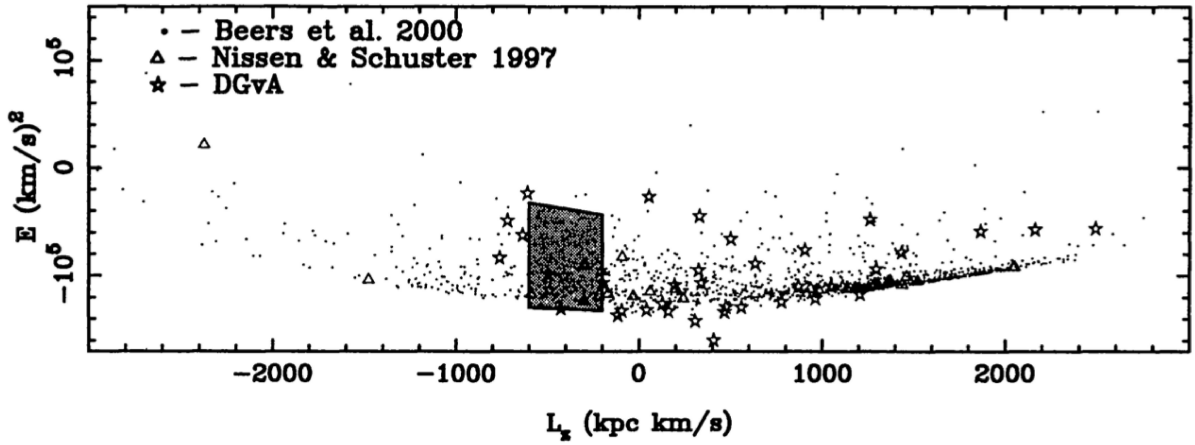}
\caption{Orbital parameters for various studies of metal-poor stars
in the solar neighbourhood.  The candidate stars from $\omega$ Cen
are chosen to lie in the shaded zone in the E-L$_{z}$ plane. (Taken
from Dinescu, 2002.)\label{dinescu}} 
\end{figure}

\citet{dinescu02} also provided a theoretical prediction of where the
$\omega$ Cen group would appear kinematically using the metal-poor star sample of Beers et al. (2000).  Figure \ref{dinescu}
shows the angular momentum -- energy plane, in which the shaded
zone represents the area where expected $\omega$ Cen candidate stars
would lie.  Dinescu defines this region by assessing three globular
clusters ($\omega$ Cen, NGC 362 and NGC 6779) that are believed to
have come from the same original parent galaxy, and argues that
their distribution may define the area in which further candidate
remnants could lie.  This region covers a large interval of energy
over an angular momentum range from about L$_{z}$= -200 to -600 kpc
km s$^{-1}$.

We have recently investigated the chemical properties of stars in
the Kapteyn moving group, first introduced by Eggen (1962). The
stars of the Kapteyn group, as most recently tabulated by Eggen
(1996), are mostly metal-poor and in retrograde galactic orbits,
so they were identified as a halo moving group.  Moving stellar
groups can originate in several ways.  Some form from a common gas
cloud. As the resulting cluster disperses, its stars dissolve into
the Galactic background yet maintain some common kinematical identity
which may be used to identify members of a particular stellar group.
Such moving groups represent a transition between bound clusters
and field stars, and are probably chemically homogeneous (see
\citet{de-silva07} and references therein). Other moving groups
appear to result from resonances in the galactic disk (eg \citet{Dehnen98}), and others possibly as the debris of accreted
and disrupted dwarf galaxies \citep{Navarro04}.  We
were interested to see whether the Kapteyn group members were
chemically similar, because this would be a pointer to the group's
origin.  It turned out that some of the group stars show chemical
peculiarities similar to those seen in $\omega$ Cen, and we will
argue that the Kapteyn group may be part of the $\omega$ Cen debris.

Concentrations of metal-poor stars at L$_z$ values near those of
$\omega$ Cen and the Kapteyn group have appeared in several recent
studies.  \citet{Dettbarn07} investigated substructure
in a sample of stars with [Fe/H]$ < -1$ from the Beers et al (2000)
catalog and identified a structure (their feature K) which appears
to be related to the Kapteyn star group.  The study of 246 metal-poor
stars with accurate kinematics by \citet{Morrison09}
also showed a substantial number of stars at weakly retrograde
values of L$_z$, similar to that of $\omega$ Cen. \citet{Klement09} find evidence of Kapteyn stream stars in their study of
halo streams from the SDSS DR7.

This paper presents a kinematic and chemical analysis of 17 members
of the Kapteyn moving group.  Section \ref{kins} presents a kinematic
analysis of the group.  In Section \ref{obs} the observations and
reduction procedures are outlined, and Section \ref{abunds} gives
the details of the chemical abundance study.  The summary and
conclusions are given in Section \ref{sumandconc}.

\section{Kinematic Analysis}\label{kins}

The initial 17 stars for this study were taken from \citet{Eggen96}, which
presents 33 Kapteyn group members.  All non-binary southern sky
members of the Kapteyn group were observed, except for the hottest
candidates, defined as (B-V)$_{\textrm{Eggen}}$ $<$ 0.40, which would have had few lines for analysis and the
coolest stars, defined as (B-V)$_{\textrm{Eggen}}$ $>$ 0.95, in which molecular features makes abundance analysis
difficult.

Table \ref{kinematics} shows the basic kinematic data for each star.
The proper motions are taken from the Hipparcos catalogue, radial
velocities are taken from various sources (see notes to Table
\ref{kinematics}) and distances are taken from the Beers et al.
(2000) catalogue.  Three stars in this sample  (BD -13 3834, G 18-54
and G 24-3) were not in the Beers et al. catalogue.  For two of these
stars, we used the method of Beers et al. 2000 to obtain distances
consistent with those in the catalogue. The distances for G 18-54
and G 24-3 were found using 

\begin{center}$M_{v} = -2.1576 + 19.9889(B-V) -15.0820(B-V)^2 +
4.7913(B-V)^3$\end{center}

taken from Beers et al.
(2000), using (B-V)=0.47 and 0.45 respectively \citep{Eggen96}.  This equation holds for all dwarf stars and was used to correspond to the dwarf gravities found for these stars via spectroscopic analysis (see Section \ref{abunds} for details).  The star BD -13 3834 was discarded from the sample due to the inability to obtain a reliable model atmosphere for this star (discussed further in Section \ref{abunds}).  This, in turn, meant a reliable distance could not be found for this star, rendering any kinematics obtained for this object essentially meaningless.  The remainder of the analysis was undertaken on only 16 stars.  

In order to study the kinematics of this moving group, space motions
were calculated using the parameters listed in Table \ref{kinematics}.
The U velocity is defined as positive toward the Galactic anti-center.
The U, V and W velocities were corrected for the solar motion of
(U, V, W)$_{sun}$ = (-10.0, +5.2, +7.2 km s$^{-1}$) from
\citet{Dehnen98}, and are shown in Table \ref{uvwlze}.    For consistency with the work of \citet{Dinescu99a}, the energy and
angular momentum were calculated using their adopted potential.
This is based on the \citet{Paczynski90} Galactic model with an
additional offset of $\Phi_o$=-12.3 x 10$^4$, introduced in order
to set the escape velocity at R = 8.0 kpc to be 500 km s$^{-1}$.
The potentials used for the bulge, disk and dark halo are shown in
Equations 3, 4 and 5 respectively.

\begin{equation}
\Phi_b = \frac{-G M_{b}}{\sqrt{R^2 + (a_b + \sqrt{z^2 + {b_b}^2})^2}}  
\end{equation}
\begin{equation}
\Phi_d = \frac{-G M_{d}}{\sqrt{R^2 + (a_d + \sqrt{z^2 + {b_d}^2})^2}}  
\end{equation}
\begin{equation}
\Phi_h = \frac{G M_h}{d} \left[\frac{1}{2} ln \left(1 + \frac{r^2}{d^2}\right)+\frac{d}{r}arctan\frac{r}{d} \right] 
\end{equation}

The bulge and disk are modeled as modified Plummer potentials,
as introduced by \citet{Miyamoto75}, while the dark halo is modeled
as a logarithmic potential.  The constants used in each Galactic
model potential are given in Table \ref{constants}.  The resulting energy E and angular momentum L$_{z}$ for each
star in the sample are also given in Table \ref{uvwlze}. The 
Lindblad diagram (E vs L$_{z}$) was constructed, including our stars,
stars from the Beers et al. catalog and the cluster $\omega$
Cen itself (see Table 4), to investigate whether there is any
kinematic connection between $\omega$ Cen and the Kapteyn group.
It was already evident that the E,L$_{z}$ values for the Kapteyn
group stars lie in the same retrograde region of the Lindblad diagram
as $\omega$ Cen.  The errors on E and L$_{z}$ will be highly correlated for some
stars, because both are dependent on distance. We ran Monte Carlo
simulations for each star to estimate the error distributions of
the derived  E, L$_{z}$ values.  Errors adopted for V$_{rad}$,
$\mu_{\alpha}$ and $\mu_{\delta}$ were those quoted in the various
sources for each star.  Errors on distance were set at 20$\%$,
consistent with the error estimates on distance in the Beers et al.
(2000) catalog.  For each star, 1000 values were drawn from Gaussian
probability distributions for V$_{rad}$, $\mu_{\alpha}$, $\mu_{\delta}$
and distance $d$, with $\sigma$ values for each equal to the errors
discussed above.  These values were then propagated through the
calculation of U, V, W, E and L$_{z}$ to obtain a probability
distribution of these derived quantities for each star.  These
probability distributions show how tightly confined the stars are
on the Lindblad diagram and, consequently, how likely the stars are
to be kinematically related to $\omega$ Cen.

\subsection{Kinematics Results}

Table \ref{uvwlze} gives the space velocities U,V,W and their error,
the angular momentum L$_{z}$ and the energy E for all stars in our
sample.  These stars were originally selected by Eggen (1996) because
of their tightly confined range of V velocities, lying between $
V=-275$ to $-300$ km s$^{-1}$.  However, with modern parallaxes,
radial velocities and proper motions, we find that Eggen's Kapteyn
group members in our study are still mostly in retrograde orbits
but occupy a more extended area of phase-space than he found, with
V values lying in a much larger range between $V=-115$ to $-368$
km s$^{-1}$.  All stars in the sample show U values well within the range observed
for the $\omega$ Cen candidate stars in the Meza et al. (2005)
study.  Meza et al found that all candidates had U values between
-300 and 300 km s$^{-1}$ and a velocity dispersion greater than 200
km s$^{-1}$.  The 16 stars in this sample of the Kapteyn group have
U values between -177 and 172 km s$^{-1}$.

\begin{figure}
\plotone{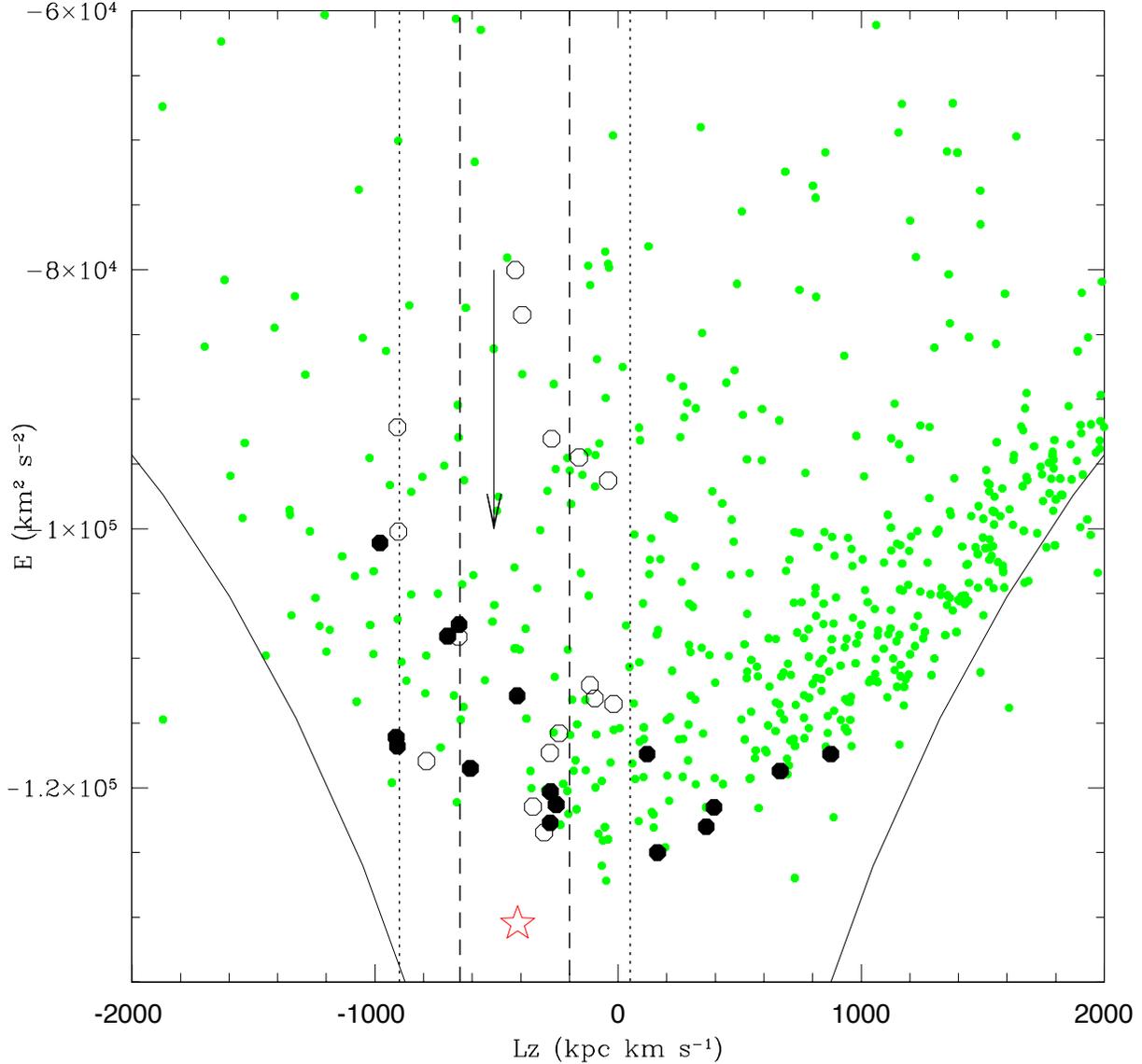}
\caption{Lindblad diagram for Kapteyn group.  The black filled circles
are the 16 Kapteyn values, as shown in Table \ref{uvwlze}, and the
red star indicates the position of $\omega$ Centauri.  The black open circles are stars from the Gratton $\omega$ Cen peak,
as seen in Figure \ref{mezafig}, with a filter of L$_z$ = -1000 to
0 kpc km s$^{-1}$.  The small green filled circles are stars from
the Beers et al. 2000 catalogue with a filter of [Fe/H] = -2.5 to
-0.5, the accepted range observed in $\omega$ Centauri.  The box
outlined by the dashed lines shows the zone in which stars are
considered to be candidates from $\omega$ Cen's host galaxy, as
defined by Dinescu (2002).  The smaller dotted lines show this same
region when taking into account uncertainties on the values of the
Local Standard of Rest (for more detail, see text).  The arrow 
indicating L$_z = -510$ kpc km s$^{-1}$ shows the L$_z$ value for
Kapteyn's star itself, using the original values from Eggen (1996).
The solid curves represent the circular orbit loci for this
potential.
\label{Lindblad}}
\end{figure}

\begin{figure}
\plotone{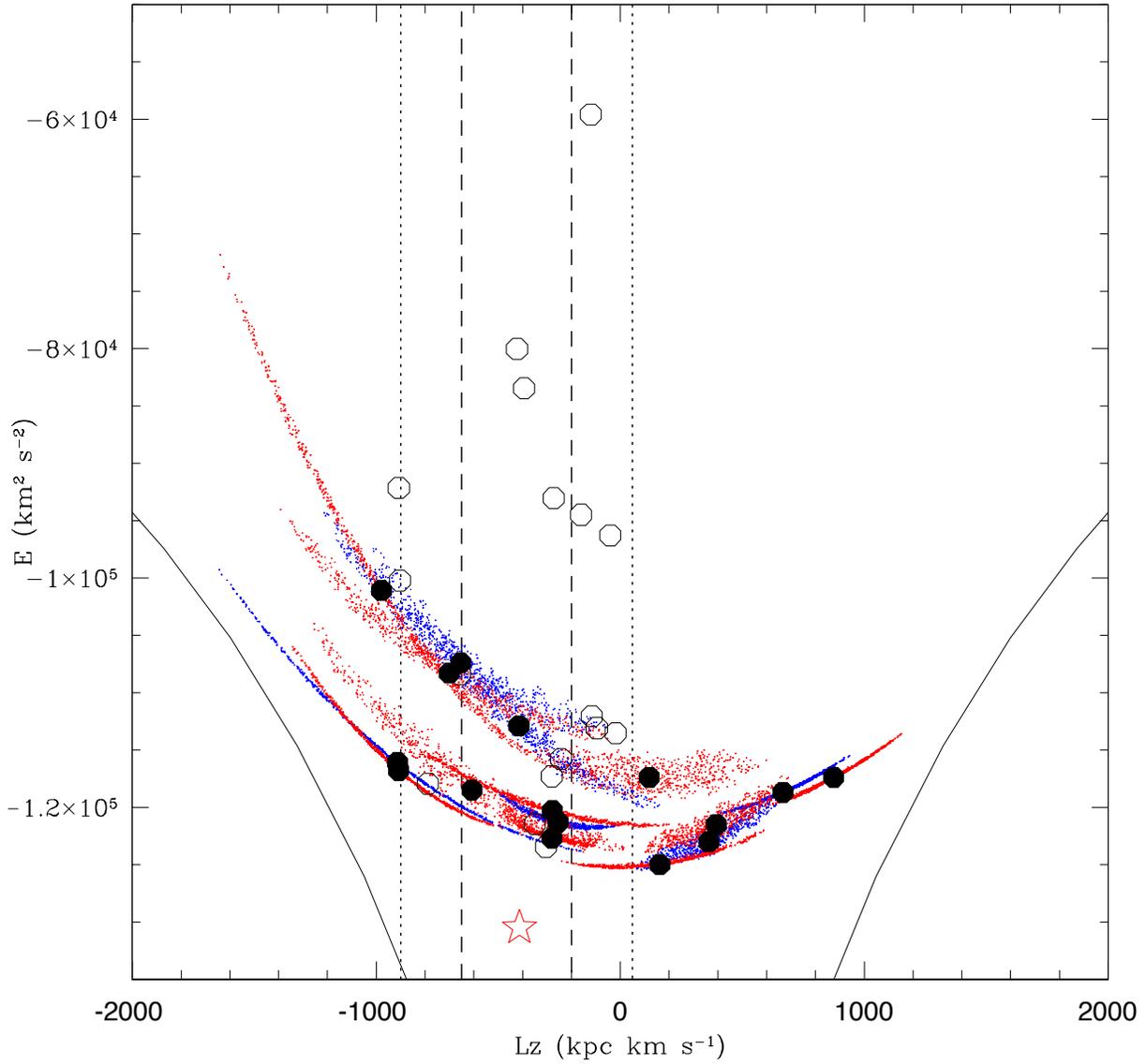}
\caption{Lindblad diagram for Kapteyn group and Gratton peak. Symbols
for the Kapteyn group stars and the Gratton peak stars, and the
vertical lines, are as in Figure 3. The smaller dots represent the
error distribution in E,L$_z$ for the Kapteyn group stars, as discussed in the text. The different
colors used are intended only to provide clarity between the
distributions for different stars.
\label{Lindbladwmc}} 
\end{figure}

The Lindblad diagram is shown in Figure \ref{Lindblad} for the
Kapteyn group stars and several other samples.  The large solid
circles are for the retrograde Kapteyn group stars (given in Table
\ref{uvwlze}).  These stars appear again in Figure \ref{Lindbladwmc}: the elongated
cloud of smaller coloured points around each solid circle show the
outcome of the Monte Carlo simulations discussed above. The points
outline the 1$-\sigma$ or 68\% confidence level for each Kapteyn
group star.  The clouds of Monte Carlo points are plotted in different
colors to provide some clarity between the different stars and their
uncertainties.  As expected, the errors in L$_{z}$ and E are highly
correlated.  The open red star is for $\omega$ Cen, using the
parameters given in Table \ref{omega}.  The total uncertainties
from a Monte Carlo simulation using  errors in V$_{rad}$, $\mu_{\alpha}$,
$\mu_{\delta}$ and distance for Omega Cen's parameters are smaller than
the starred symbol.  The smaller green circles are stars from the
Beers et al. catalogue, with an [Fe/H] filter applied to cover $-2.5
<$ [Fe/H $< -0.5$, the abundance range observed in $\omega$ Cen.  The region in which Dinescu (2002) suggested that candidate stars
from Omega Cen's host galaxy could lie is outlined by the dashed
line (discussed in more detail in Section 1).  This range depends
on the adopted velocity of the Local Standard of Rest (taken here
to be 220 km s$^{-1}$. The dotted lines indicate the Dinescu range
using extreme values of the LSR velocity of $190$ km s$^{-1}$ \citep{Olling00} and $250$ km s$^{-1}$ \citep{Reid09}.  The
arrow indicating L$_z = -510$ kpc km s$^{-1}$ is the L$_z$ value of
Kapteyn's star using original values from Eggen (1996).  The solid
curves near the bottom of the figure show the prograde and retrograde
circular orbit loci for the adopted potential.

Figure \ref{Lindblad} shows many prograde field stars from the Beers
catalogue, because the Beers sample includes a lot of thick disk
stars which are primarily in prograde orbits.  $\omega$ Cen stands
out as the highly bound object at L$_{z}=-413$ kpc km s$^{-1}$ and
E$ =-1.30 \times 10^{5}$ km$^{2}$s$^{-2}$, consistent with the values
published by Dinescu et al. (1999) of L$_{z}=-406$ kpc km s$^{-1}$
and E$ =-1.31 \times 10^{5}$ km$^{2} $s$^{-2}$.  The Kapteyn group stars in Figure 4 appear to lie mainly in two
bands, with a third small concentration of stars near (L$_z$,E)$ =
-200$ kpc km s$^{-1}$, $ -0.95 \times 10^5$ km$^{2}$s$^{-2}$, defined
by three of the Gratton peak stars and a number of the Beers stars
(see Figure 3). The appearance of these bands is accentuated by the
correlated errors on E and L$_z$, and will be discussed further in
Section 5.  Figure \ref{Lindblad} shows that nine of the Kapteyn stars lie within
the box covering the Dinescu et al. region of $\omega$
Cen candidates.  Within the 1-$\sigma$ uncertainties, this number
increases to 12 members of the Kapteyn group.  There are four
prograde stars in the Kaptyen sample. Two of these stars lie such
that the ends of their 1-$\sigma$ distribution fall near the
extreme of the $\omega$ Cen region, leaving only two stars
(HD13979 and CD -30 1121) which are clearly separate from the
$\omega$ Cen candidate region.

In the following discussion, the 16 Kapteyn stars are partitioned
into two groups. From Figure 4, fourteen stars are taken as kinematically
associated with $\omega$ Cen and are displayed as red filled squares,
while two of the stars are likely to have no kinematic connection
to $\omega$ Cen debris and are plotted as black open circles.

\section{Observations and Reduction}\label{obs}

Observations were taken with the echelle spectrograph on the 2.3m
telescope at Siding Springs Observatory, Australia, over the period
2006 July to 2007 August.  Complete wavelength coverage from 4120
to 6920\AA\ was achieved, with a resolution of about 25,000 and
an average signal to noise of per resolution element of about 70.
Bias frames, quartz lamp exposures for flat-fielding and Th-Ar
exposures for wavelength-calibration were all included in the
observing runs.  Radial velocity standards were also observed at
the beginning and end of each night.  The data were reduced using
standard IRAF routines in the packages \textbf{imred}, \textbf{echelle} and \textbf{ccdred}
with no deviation from the normal reduction procedures.

\section{Abundance Analysis}\label{abunds}

Both photometric and spectroscopic atmospheric parameters were
derived for each star.  Photometric temperatures were derived via
the Alonso (1999) calibration with V-K  using V magnitudes from
Eggen (1996) and K magnitudes from the 2MASS
catalogue\footnote{http://irsa.ipac.caltech.edu}.  Photometric
gravities were found via absolute magnitudes and bolometric correction
calibrations provided by \citet{Alonso96b} and \citet{Alonso99}
using the standard relationship shown in Equation 6 and assuming a
mass for these stars of 0.8M$_{\odot}$.

\begin{equation} 
log(g) = 0.4 (M_v + BC_v - 4.75) + 4.44 + 4 log\left(\frac{T_{\textrm{eff}}}{5770}\right) + log(0.8)
\end{equation}

\begin{figure}
\plotone{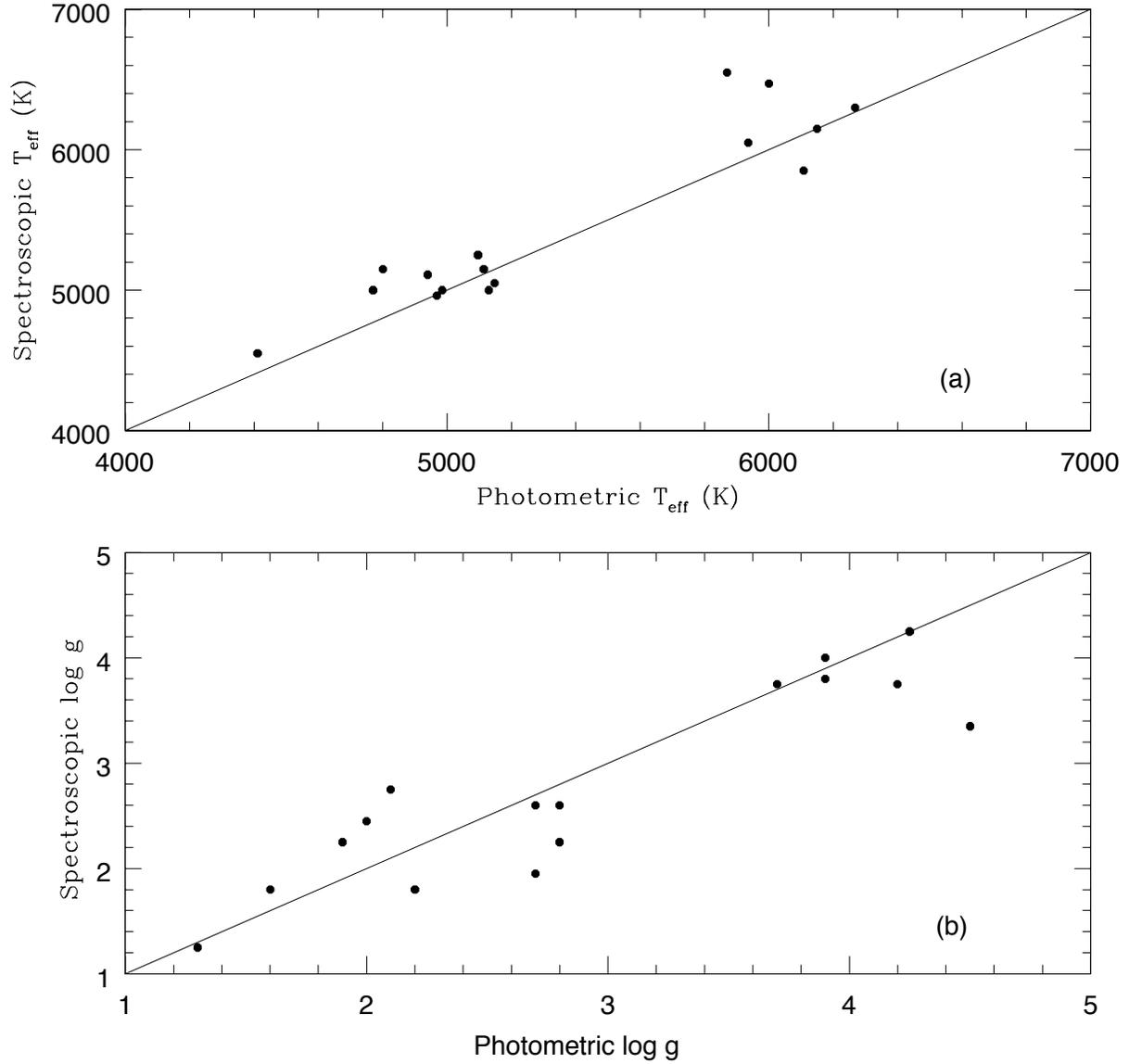}
\caption{Comparison of a) photometric and spectroscopic temperatures
and b) photometric and spectroscopic gravities.  In each plot, the one-to-one relation is overplotted.
\label{spvscs}} 
\end{figure}

Measured equivalent widths of both Fe I and Fe II lines were used
to derive spectroscopic temperatures and gravities.  The
usual comparison between excitation potential and derived abundance
was made to determine effective temperature while the balance between
abundances derived from neutral and ionised iron lines was used to
refine the gravity of the star.  The microturbulent velocity was
obtained by measuring Fe I lines and ensuring that the derived
abundances be independent of equivalent width.  Figure \ref{spvscs}
shows the agreement between spectroscopic and photometric temperatures
derived.  

With two exceptions,  the photometric and spectroscopic temperatures
differed by less than 200K, while photometric and spectroscopic gravities
differed by less than 0.50 dex.  Atmospheric parameters obtained
spectroscopically, which were used throughout this analysis, are
shown in Table \ref{atmosparam} for each of the sample stars.  The
atmospheric parameters derived via spectroscopic analysis were
adopted instead of the photometric ones for consistency, because
the abundance analysis uses spectroscopic data.  In the case of one star, BD -13 3834, a reliable model atmosphere could not be found.  While the spectroscopic and photometric effective temperatures agreed to within 250K, the disagreement between the log g values was 1.3 dex.  This was considered too large a disagreement to make either value reliable and so this star was discarded from the sample.  

The estimated errors on T$_{\textrm{eff}}$ are taken from the largest
difference found between spectroscopic and photometric values.  The
errors on log g are found from either the largest difference between
spectroscopic and photometric values or the uncertainties on the
observables of M$_v$ and BC$_v$, whichever is larger.  The errors
on $\chi$ are taken to be constant for all stars.  Typically,
the errors are about $\Delta$T$_{\textrm{eff}}$=150K, $\Delta$ log g =
0.2 dex and $\Delta \chi$ = 0.25 km s$^{-1}$.  The effect of these
uncertainties on derived abundances is discussed further in Section
\ref{uncabund}.  All abundances were obtained via the method of spectrum synthesis
using the MOOG code \citep{Sneden73} using \citet{Anders89} solar abundances.  The analysis for these stars
included atomic line lists from
 Kurucz\footnote{http://kurucz.harvard.edu} with refined oscillator
strengths for all lines via a reverse solar analysis using the Kitt Peak solar spectrum \footnote{http://bass2000.obspm.fr/solar$_{-}$spect.php}.  The Na D lines were used to obtain the sodium abundance.  In metal-poors stars, such as these, these lines do not saturate and provide a reliable estimate of the sodium abundance.  The spectrum synthesis program used (MOOG) has an option to deal with strong lines separately during the analysis.  Hyperfine splitting components were taken into account for the Na D lines and lines of both Cu and Ba.  This
reverse solar analysis was done with a solar model of T$_{\textrm{eff}}$=5770K
and log g=4.44.

\subsection{Uncertainties on derived abundances}\label{uncabund}

As well as error introduced by uncertainty in continuum placement, which is always minimised as much as possible, there are three main sources of uncertainty in these derived
abundances.  The largest source of uncertainty is the spread in
derived abundances from multiple lines of the same element.  Although
in theory this should be small, different lines of the same element 
can give abundances differing by as much as 0.2 dex.  This can be
due to incorrect log gf values, although in this study a reverse
analysis using the sun was done for all elements to reduce this
problem (see earlier in Section 4 for more details).  Non-LTE
effects, blending and data quality problems can also contribute. 
This dominant error contribution is tabulated as the $\sigma$ value
in Table 7. Where abundances are obtained from only one line (as 
for Zr and occasionally other elements), the $\sigma$ value is the
uncertainty from a single line measurement, as discussed below.  The second source of uncertainty is the accuracy with which is is
possible to determine an abundance from any particular line. This
uncertainty can vary depending on line strength, blending and
crowding in the region, and the overall quality of the data. In
this study, this error contribution is estimated to be no more than
about 0.1 dex.  The third source of uncertainty is the difference in abundances
obtained using different atmospheric parameters in the stellar
models.  This depends on how well the atmospheric parameters
can be constrained.  Table \ref{abundsens} shows the dependence of
derived abundances on the chosen atmospheric parameters.  These
values were obtained by choosing equivalent width values which gave
abundances representative of those obtained via spectrum synthesis
and then running an abundance analysis on these equivalent widths
with four different models, with $\Delta$T$_{\textrm{eff}}$=-200K, $\Delta$log
g=-0.3 and $\Delta$$\nu$=-0.30km s$^{-1}$.  These $\Delta$ values
represent slightly higher uncertainties on the atmospheric parameters
than the ones estimated in this study (see Table \ref{atmosparam}).
It can be seen that atmospheric parameters influence different
lines, stressing the importance of ascertaining the correct model
atmosphere at the outset.  

\subsection{Abundance Results}\label{abresults}

Four stars in the current sample have been previously analysed for abundances by studies cited in this paper.    A brief summary of agreement is given here as a consistency check.  One star in our sample, HD 181743, is in common with the Gratton et al.
(2003) sample. The effective temperature differed by only 82K, but
the log g values showed a larger difference of 1.05. All abundance
values, including [Fe/H], differed by less than 0.2 dex, within 
the quoted errors of this study and the Gratton et al study.
Another of the stars in the Kapteyn group, HD 111721, was previously studied by \citet{Fulbright00}, and is included in the list of stars compiled by \citet{Venn04}.  For this star the differences in adopted atmospheric model between this study and \citet{Fulbright00} was only 175K in effective temperature, 0.3 in gravity and 0.06 in [Fe/H].  Within uncertainty, all abundance determinations for Na, Mg, Ca, Zr and Ba for this star agreed with Fulbright's values, with no value differing by more than 0.15 dex.  Two more stars, HD 13979 and HD 186478, were also studied by \citet{Burris00}, which is again included in the \citet{Venn04} compilation, and abundances for the s-process elements were derived.  The abundances obtained for Zr and Ba in this study agree with those of \citet{Burris00} to within 0.02 for HD 186478, although the discrepancies are larger for HD 13979, with differences around 0.3 dex.  A reason for this is not obvious, as the atmospheric model adopted differed by only 25K in effective temperature and 0.2 in gravity.

\begin{figure}
\plotone{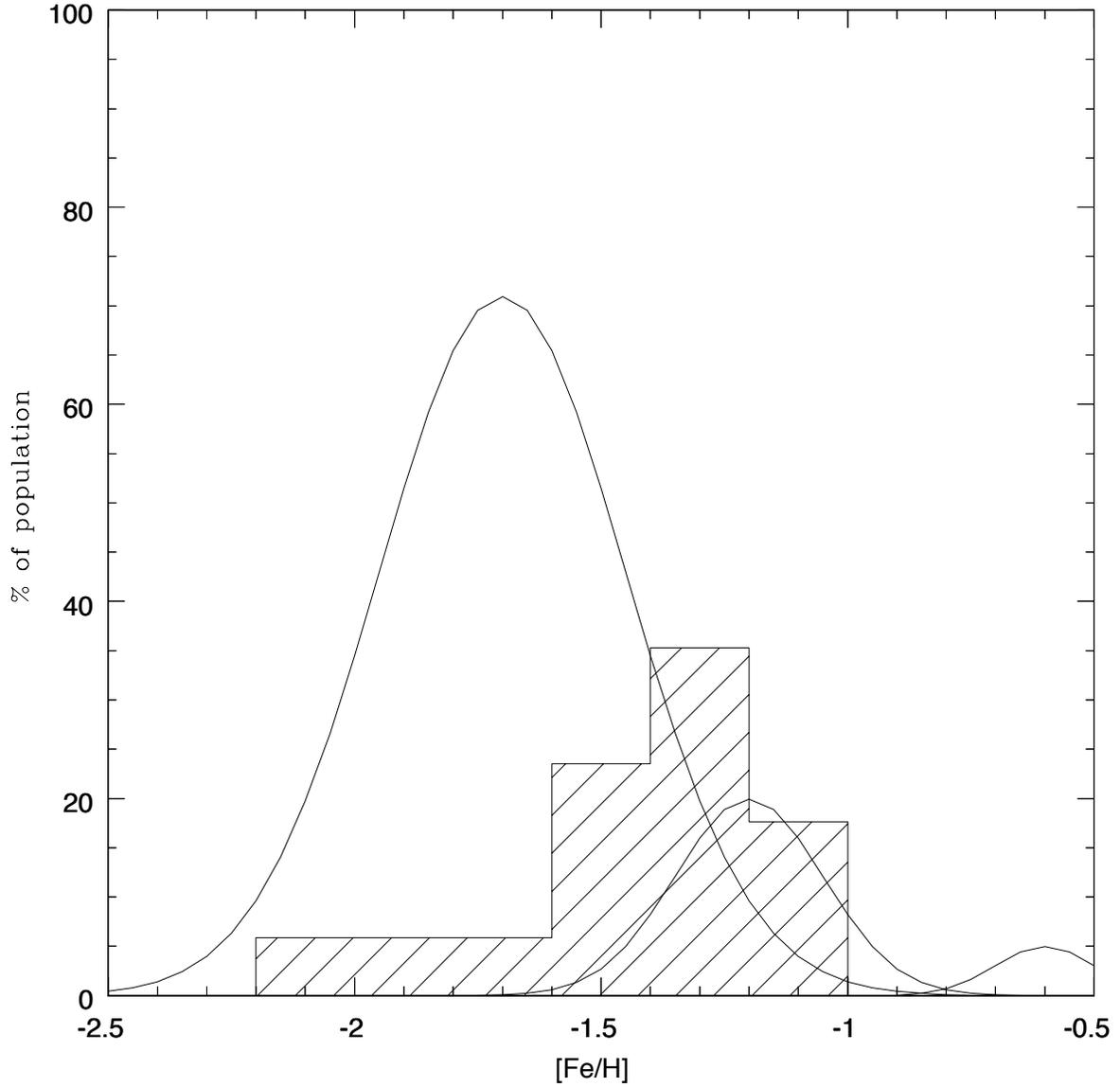}
\caption{Histograms of [Fe/H] for the sample of 17 Kapteyn group
stars, 15 of which may be kinematically related to $\omega$
Cen.  The three gaussian curves represent the three suggested
populations of $\omega$ Centauri at [Fe/H]=-1.7, -1.2 and -0.6,
with the percentage of stars observed at that metallicity shown as
70\%, 20\% and 5\% respectively (Norris et al. 1996)
\label{fehisto}}
\end{figure}

Figure \ref{fehisto} shows a histogram of the [Fe/H]
values of all 16 of the Kapteyn group stars.  Also shown are the
three suggested sub-populations of $\omega$ Centauri at [Fe/H] = -1.7,
-1.2 and -0.5 \citep{Norris95}.  The metallicity distribution of the
stars in the Kapteyn group falls towards the upper end of the extended
MDF for the $\omega$ Cen stars. but is consistent with the cluster MDF.  Abundances obtained for Na, Mg, Ca, Cu, Zr and Ba and their
uncertainties are given in Table \ref{abundances}.  The errors
quoted are either the spread in derived abundances from different
lines, or set to a standard value of 0.1 dex, representing the
uncertainty in any individual measurement, whichever is the larger.

\begin{figure}
\plotone{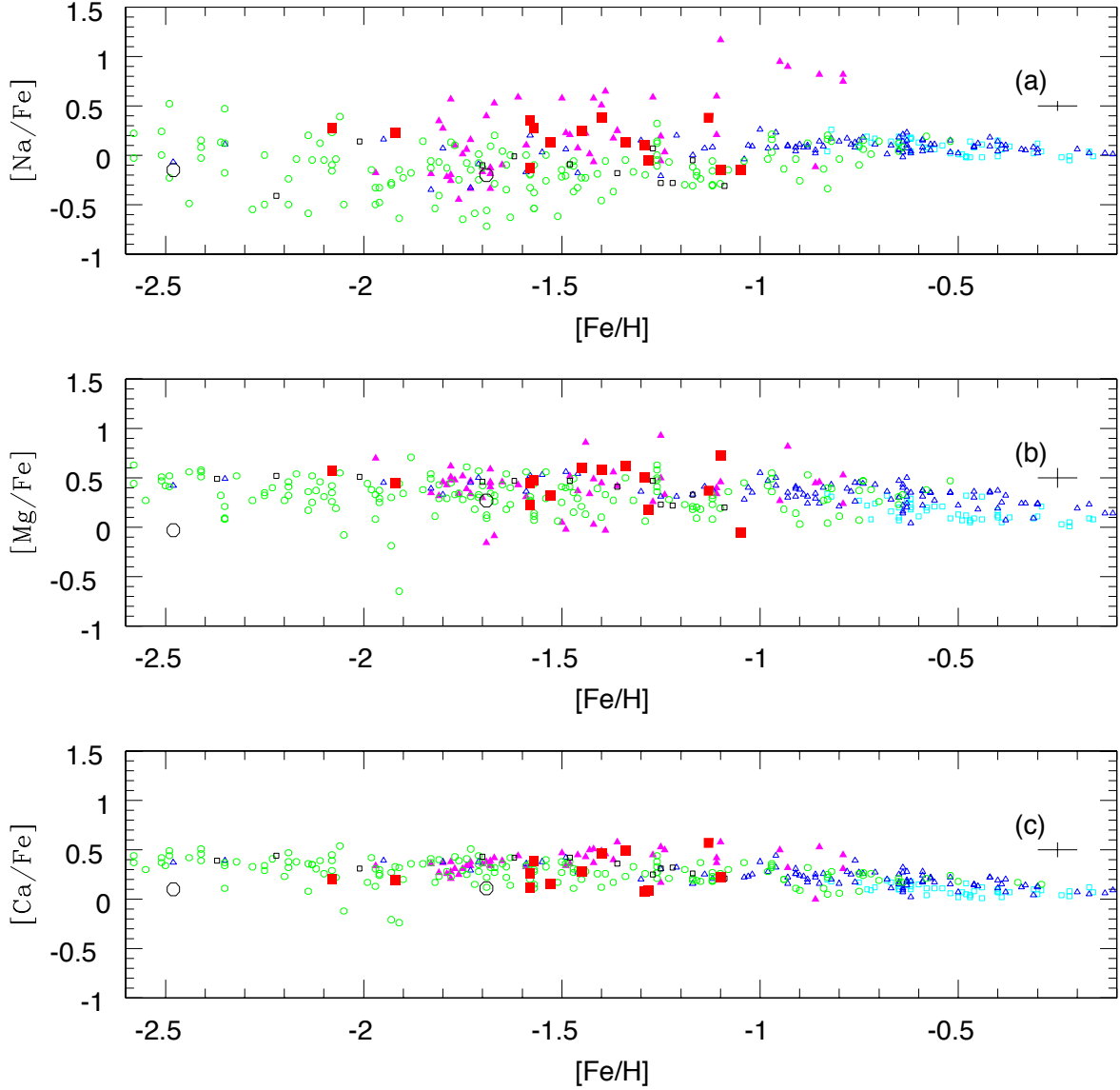}
\caption{Plots of a) Na abundance, b) Mg abundance and c) Ca
abundance.  The red filled squares are those stars that may be
kinematically linked to $\omega$ Cen, while the open black circles
are those that do not appear to have kinematics link to the globular
cluster.  The small open black squares are stars from the Gratton
peak (see Figure \ref{mezafig}).  Also shown are stars from the
thin disk (small open cyan squares), thick disk (small open blue
triangles) and halo (small open green circles) from Venn et al.
2004; stars from $\omega$ Cen (small filled magenta triangles) from Norris 1995 and Smith
2000.   Typical error bars are shown in the top right of each
plot.
\label{light}} 
\end{figure}

\begin{figure}
\plotone{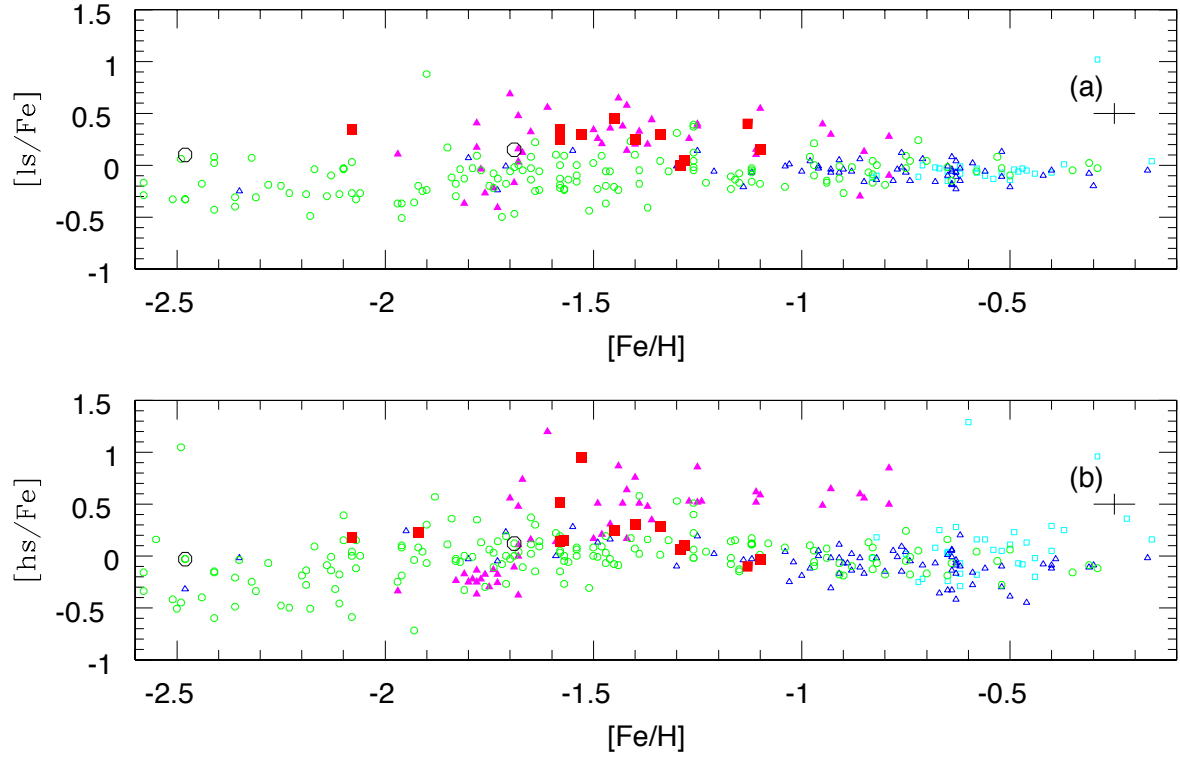}
\caption{Plots of a) light s-process element abundances and b) heavy
s-process element abundances.  Symbols are the same as in Figure
\ref{light}.  The elements plotted in a) are Zr for the Kapteyn
group and $\omega$ Cen stars and Y for the thin disk, thick disk
and halo stars.  The heavy s-process element plotted in b) is Ba
for all samples.   Typical error bars are shown in the top right
of each plot.
\label{heavy}} 
\end{figure}

Figure \ref{light} shows the abundances for Na, Mg and Ca while
Figure \ref{heavy} shows the abundances for the light and
heavy s-process elements.  In both figures the red filled squares
are those stars that are likely to be kinematically linked to
$\omega$ Cen, while the black open circles are those that are
probably not kinematically associated with the cluster.  Also shown
are stars from \citet{Venn04} for the thin disk (cyan open squares),
thick disk (blue open triangles) and halo (green open circles).
Stars in $\omega$ Cen itself, from studies by \citet{Norris95} and
\citet{Smith00} are shown as magenta filled triangles. The Gratton
peak stars are shown as small open black squares.

First we compare the abundances in $\omega$ Cen and in the field
halo stars. Then we compare the Kapteyn group abundances with those
in the cluster and the field halo, in order to ascertain whether
the Kapteyn group stars show the abundance patterns of either
of these two populations.  The Na abundance in Figure \ref{light}a show a clear separation
between the $\omega$ Cen studies (magenta filled triangles) and the
halo stars (green open circles), although this distinction is not
so clear for Mg.  For Ca, there is separation in the more metal-rich
regime ([Fe/H]$>-1.5$) with the distinction becoming less for more
metal-poor stars.  In Figure \ref{heavy}a the light s-process elements show show a
marked difference between the abundances observed in $\omega$ Cen
and the field halo stars, with $\omega$ Cen showing a strong relative
enhancement in the light s-process elements.  In Figure \ref{heavy}b,
an even larger difference in Ba abundance is evident, with $\omega$
Cen showing a spread in Ba abundance from [Ba/Fe]=+1.0 to -0.4 over a range of metallicities .  Again, as for Ca,
this distinction becomes less at [Fe/H] near -1.7, where the
separation is less and the two populations appear to overlap more.  From these two figures it seems clear that there is a distinct
difference between the Kapteyn stars believed to be kinematically
related to $\omega$ Cen (red filled squares) and those that appear
to be prograde field stars (black open circles).  The prograde
members of the Kapteyn sample show abundances that are consistent
with being part of either the thick disk or halo populations.

The main point of interest here is that the Kapteyn stars which
appear to be kinematically related to $\omega$ Cen in Figure
\ref{Lindblad} (red filled squares) share distinct abundance
similarities with those stars from previous $\omega$ Cen studies
(magenta filled triangles).  The abundances for the potential
$\omega$ Cen candidates among the Kapteyn stars could be drawn from
the same abundance distribution as the $\omega$ Cen stars, although
we note that there is much more scatter in the abundance distributions
for most elements in the $\omega$ Cen stars.  This larger scatter
may be several causes: difference in the two studies making up this
sample (\citet{Norris95} and \citet{Smith00}); a genuine scatter
in the abundances observed in $\omega$ Cen stars; or a combination
of a genuine scatter and observational uncertainties.  Whatever the reason for this large scatter in the cluster star
abundances, it seems clear that the Kapteyn candidates show abundance
anomalies relative to the field halo which are similar to those
shown by the stars of $\omega$ Cen.  The similarities between the
Kapteyn candidates and the $\omega$ Cen stars, relative to the field
halo, are clearest in Na, Mg and the s-elements.

Although the copper abundance was derived for only five of the stars
in our sample, the results are worth mentioning. $\omega$ Cen is known
for its unique copper signature. At a constant value of [Cu/Fe] $\sim
-0.6$, it is distinct from the halo stars in the same metallicity
range, which show an increase with metallicity up to [Cu/Fe] = -0.3
as discussed by Smith et al. (2000).  Of the four stars for which
copper was measured in our sample, all show deficiencies, with 
[Cu/Fe] values between -0.47 and -0.60.  This adds weight to the
case that this group of stars come from the same population as
$\omega$ Cen.  Copper was also detected in a fifth star, CD -30.1121, which is not believed to be kinematically linked to $\omega$ Cen.  The copper value in this star was found to be [Cu/Fe] = -0.7.  Copper has been seen to decrease sharply with metallicity among field halo stars, as shown in \citet{Sneden91}.  With a metallicity of [Fe/H] = -1.69, the copper value seen in CD -30.1121 is consistent with the field halo star trend seen in copper and the observed low copper does not necessarily exclude it as a field halo star, as the kinematics suggest.  

\subsection{Statistical Analysis} 

Statistical tests were undertaken on the abundance distributions to determine the
likelihood that the three groups of stars (Kapteyn group, $\omega$
Cen and the halo) come from the same population. The $\omega$ Cen group consisted of stars from the \citet{Norris95} and \citet{Smith00} studies and includes the stars from \citet{Gratton03} only for elements in common with this study, Na, Mg and Ca.  The halo group is taken from the \citet{Venn04} compilation and is defined as being any halo population star that falls within the plausible range of [Fe/H] values for $\omega$ Cen as seen in Figure \ref{fehisto}, that is [Fe/H] = -0.5 to -2.5.  However, for completion, halo stars outside these values, while not included in the statistical analysis, are included in the abundance plots.  Table \ref{meansstds} shows the
mean, standard deviation $\sigma$ and standard deviation of the
mean $\sigma_{mean}$, as defined by $\sigma_{mean}$=$\sigma$/$\sqrt{N}$ where N is the number of stars in each sample, also given in Table \ref{meansstds}. Figure 9 shows this information graphically for
the five elements studied, with the Kapteyn group again shown as filled 
red squares, the $\omega$ Cen stars as filled magenta triangles and the
field halo stars as open green circles. It is evident from this figure 
that in all elements, except Ca and to a lesser extent Mg, the field
halo stars are chemically distinct from $\omega$ Cen.  It is also
clear that, except for Ca, the Kapteyn group and $\omega$ Cen
populations overlap in [X/Fe].

It should be noted that results from a very recent study of 66 red giant stars in $\omega$ Cen by \citet{Johnson09} strongly supports these results.  They find a general increase in [Na/Fe] with increasing metallicity, from $<$[Na/Fe]$>$=+0.03 ($\sigma$=0.32) at the lower metallicity end to $<$[Na/Fe]$>$=+0.86 ($\sigma$=0.12) for higher metallicity.  They also find $<$[Ca/Fe]$>$=+0.36 ($\sigma$=0.09), La values varying from [La/Fe]=-0.4 to +2.0 and $<$[Eu/Fe]$>$=+0.19 ($\sigma$=0.23).  These results strongly overlap with the previous studies of $\omega$ Cen stars used as a comparison with the Kapteyn group stars in this study.

A parametric t-test and a non-parametric Kolmogorov-Smirnov
test were performed on the abundance distributions. The results are
shown in Tables 10 and 11 respectively. In these tests, a small P
value means that the two populations are distinct.
We adopted the standard
criterion that P$ < 0.05$ is needed to reject the null hypothesis.
The tables show that both the parametric and
non-parametric test gave the same results. Tests in which the P
value satisfied the requirement are shown in bold, indicating that
the two populations compared are statistically different. $\omega$ Cen
and the halo were seen to be distinct separate populations in all
of the five element distributions, in both tests. The Kapteyn
group and the halo were statistically different in [Na/Fe], [ls/Fe]
and [hs/Fe] while the [Mg/Fe] and [Ca/Fe] comparison is inconclusive.
The Kapteyn group and $\omega$ Cen differ only in [Ca/Fe].  While the [hs/Fe] abundance patterns seen in Kapteyn and $\omega$ Cen in Figure \ref{heavy}b differ slightly, there is enough overlap in the two groups for this overlap to be statistically significant and the statistical analysis suggests these two distributions can come from the same population.  There appear to be  only two Ba enhanced stars in the Kapteyn group sample, although both are seen at values consistent with Ba enhancements seen in $\omega$ Cen stars.  The apparent lack of Ba enhanced stars in the Kapteyn group may be due to a small sample size of only 16 stars.  

Together with Figures 7 and 9, these statistical tests show that [Ca/Fe] does
not differ enough between the populations to give reliable discrimination.
The [Ca/Fe] relationship in Figure 7 is tight and all populations
overlap.  However, the difference seen in the remaining four elements
provide enough information to show that the field halo stars are
different from both the Kapteyn stars and the $\omega$ Cen stars
and that the Kapteyn group stars are chemically similar to those
in $\omega$ Cen.

\section{Discussion and Conclusion}\label{sumandconc}

The kinematic and abundance analysis in this study suggests that
at least our 14 members of the Kapteyn group and potentially many
more stars in retrograde orbits which were not observed in this
study, could be remnants of tidal debris stripped from the parent
galaxy of $\omega$ Cen, or even from the cluster itself,  during its 
merger with the Galaxy. 

Our study provides the first detailed chemical evidence of field
stars that appear to be both kinematically and chemically related
to $\omega$ Cen. It may lend weight to the view that $\omega$ Cen
is the remnant nucleus of a disrupted dwarf galaxy which was accreted
by the Milky Way, by providing chemical evidence of tidal debris
among the Galactic field stars.

The three-banded structure seen in the Lindblad diagrams of Figures
3 and described in Section 2.1 may be indicative of stars shed with
different energies from different wraps of the decaying orbit of the
parent galaxy around the Milky Way. The reader is referred to Figure
6 in Meza et al. (2005) which shows several distinct E--L$_z$ curves
from their numerical simulations of merger debris.  In this study there is currently no evidence for an abundance difference between stars of different wraps of the orbit.  A study involving higher quality data and a larger number of stars from these wraps is underway.  
The presence of
this banded structure, along with the present Galactic radius of
$\omega$ Cen, suggest the original parent galaxy was relatively
massive in order for dynamical friction to have the required effect,
and was also relatively dense in order to survive the Galactic tidal
stresses in to the current orbital radius of $\omega$ Cen.

What else can we infer about the parent galaxy?  If our stars are
debris from the parent galaxy, then the galactic metallicity-luminosity
relationship and the mean metallicity of the sample ($\langle {\rm
[Fe/H]} \rangle = -1.5$) indicate that the parent galaxy's luminosity
would have been about $M_V \sim -11$.  Its luminosity would have
been about $2 \times 10^6$ L$_\odot$, and its stellar mass $\sim 4
\times 10^6$ M$_\odot$. This stellar mass is comparable to the
present stellar mass of $\omega$ Cen itself.  The total mass of the
parent galaxy could have been as high as $5 \times 10^8$ M$_\odot$
if it were a dwarf galaxy with a dark matter content similar to the
Fornax dSph (e.g \citet{Walker06}).

What are the chances of finding $\omega$ Cen debris stars in the
solar neighborhood from such a low-mass galaxy? We can use the
Meza et al. simulation as a guide, although a detailed comparison
is not appropriate as these authors note, because their end-product
galaxy is not like the Milky Way. In summary, the debris of their
satellite is well mixed azimuthally by redshift 0.48, at least
within 10 kpc radius from the center of their parent galaxy, and
is confined to a disk-like layer. The satellite is disrupted in its
last three perigalactic passages, leaving a significant amount of
substructure in E and L$_z$. We focus on the lowest $L_z$ substructure
in their simulation, as a qualitative counterpart to the proposed
$\omega$ Cen debris. Most of our stars lie in a volume of radius
$\sim 500$ pc around the sun, and in a broad range of retrograde
L$_z$. For comparison, we can estimate the fraction of the Meza et
al. debris which would lie within our $500$ pc volume and within
their lowest L$_z$ substructure: it is about $1.5 \times 10^{-4}$.
We could therefore expect to find about 700 debris stars in our
volume and within the lowest angular momentum substructure, if we
use the Meza et al.  simulations as a guide and assume that the
typical star in this population has a mass of about 0.8 M$_\odot$.
In the Dinescu region shown in Figure 3, we have compiled a total
of about 100 stars.  Even if all of them turn out to be $\omega$
Cen debris stars, then this comparison may indicate that there are
plenty more $\omega$ Cen debris stars to be found nearby.

We should also consider the possibility that the Kapteyn group stars
were stripped from the cluster itself as its outer regions were
disrupted by the Galactic tidal field. This would be the immediate
inference from the the chemical similarity which we have demonstrated
between the Kapteyn group stars and  Cen stars. This seems unlikely,
because the in-spiralling time under the influence of dynamical
friction is in excess of the Hubble time for a cluster with the
mass of $\omega$ Cen; also, we note that the Kapteyn group stars
are significantly more energetic than the cluster itself. It seems
likely that the group stars came from the body of the parent
dwarf galaxy, and therefore that the chemical peculiarities of
$\omega$ Cen are shared by its parent, and may well have originated
in gas which was funneled into the cluster, possibly over an
extended period, as suggested by \citet{Norris97}.  Detailed numerical models of the chemical and dynamical evolution of
$\omega$ Cen as its galaxy loses mass in the Galactic tidal field are
needed.

{\bf Acknowlegements}
ECW is funded by Australian Research Council grant DP0772283. KCF
acknowledges his late colleagues Olin Eggen and Alex Rodgers for
their work on the Kapteyn's star group which prompted this paper.  We would also like to add our thanks to the referee (D. Casetti-Dinescu)
for many helpful comments and suggestions.. 
We thank Mike Bessell for his major improvements to the echelle
spectrograph which made this work possible; Daniela Carollo for
discussions which led to the association of the Kapteyn's star group
with $\omega$ Cen and for comments on an earlier draft of this paper; John Norris and Tim Beers for discussions and encouragement;
and the SSO staff for maintaining the SSO instrumentation.

\clearpage

\clearpage

\clearpage

\clearpage

\clearpage

\clearpage

\clearpage
\begin{figure}
\plotone{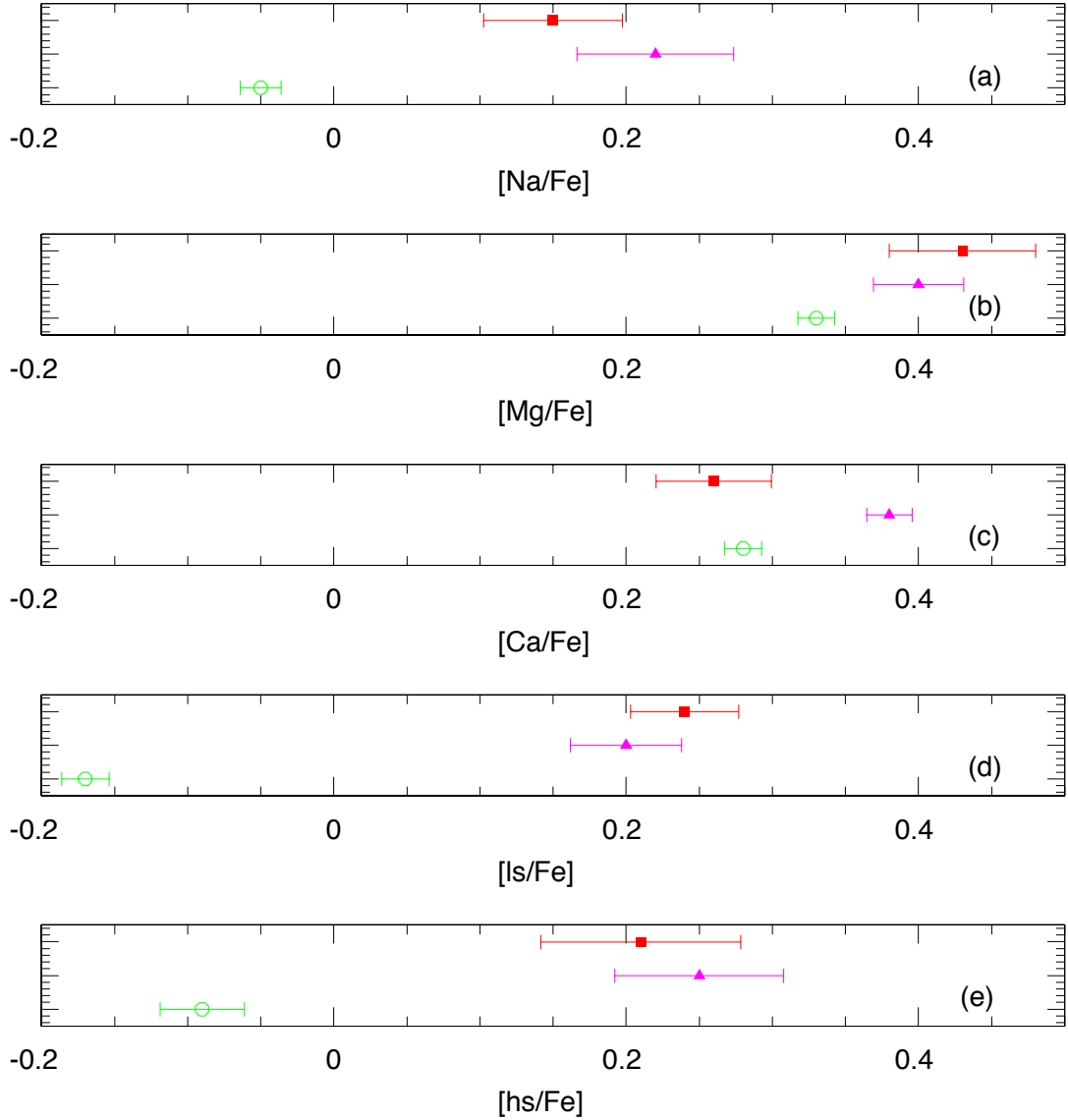} 
\caption{The mean abundance and standard deviation of the mean
abundance of the Kapteyn group stars (filled red squares), $\omega$
Cen stars (filled magenta triangles)and halo stars (open green
circles) for the five different elements studied: a) [Na/Fe], b) [Mg/Fe], c) [Ca/Fe], d) [ls/Fe] and e) [hs/Fe].  It is easy to see the mean abundance and distributions of each population and compare to the others.  The y axis simply spreads the three different groups out for clarity and ease of comparison.  
\label{means} }
\end{figure}

\clearpage

\begin{deluxetable}{lcccccc}
\tabletypesize{\scriptsize}
\tablecaption{Kinematics for the Kapteyn group of stars. \label{kinematics}}
\tablewidth{0pt}
\tablehead{
\colhead{Star} & \colhead{RA} & \colhead{DEC} & \colhead{$\mu_{\alpha}$} & \colhead{$\mu_{\delta}$} 
& \colhead{$V$$_{\textrm{rad}}$\tablenotemark{a}} & \colhead{Distance\tablenotemark{a}} \\
\colhead{} & \colhead{$^{\circ}$} & \colhead{$^{\circ}$} & \colhead{\textrm{m}$''$ \textrm{yr}$^{-1}$} & \colhead{\textrm{m}$''$ \textrm{yr}$^{-1}$} & \colhead{km s$^{-1}$} & \colhead {pc}
}
\startdata  
HD 13979 & 33.84 & -25.92 & 29.78 & -47.67 & 54.0 &453\\
HD 21022 &  50.59     & -32.99    & 31.13  &  -32.51   & 110.0    &   1115 \\
 HD 110621  & 190.93 & -44.68  &-221.96      &     -17.30      &219.0&156\\
 HD 111721  & 192.86  &-13.49  &-273.70       &    -321.99      &27.0 &162\\
 HD 181007  & 289.87  &-20.43  & -39.97       &    -170.25       &-2.0&396\\
  HD 181743  & 290.93  &-45.08  & -70.60       &    -814.46       &26.0 &92\\
 HD 186478  & 296.31  &-17.49  & -22.23       &     -84.42      &31.0 &1017\\
 HD 188031  & 298.74  &-42.65   &  2.87       &    -436.00      &-139.0 &163\\
 HD 193242  & 304.98  &-19.91 &  -20.64       &     -97.64       &-128.0 &314\\
 HD 208069  & 328.61  &-30.26   &132.14       &    -198.67      &-167.0 &285\\
  HD 215601  & 341.70  &-31.87 &   56.11       &    -148.35       &-35.0 &278\\
 HD 215801  & 342.12  &-46.06    &33.49       &    -305.71       &-86.0 &159\\
 BD -13 3834  & 212.61 & -13.93  &-337.04       &    -458.34       &126.0\tablenotemark{b}& ...\\
CD -30 1121  &  44.25&  -30.41&    31.11       &     -28.51    &104.0 &629\\
CD -62 1346  & 316.51 & -61.56  & -14.97       &    -102.19 & 127.0 &299\\
G 18-54       &337.90 &2.16 &51.69 & -328.64& -210.4\tablenotemark{c}& 178\tablenotemark{d}\\
   G 24-3  & 301.43  &  4.05  &-133.27       &    -171.24      &-208.9\tablenotemark{c} & 175\tablenotemark{d}\\ \enddata
\tablenotetext{a}{Beers et al. 2000 catalogue}
\tablenotetext{b}{Wilson 1953 catalogue}
\tablenotetext{c}{\textrm{L}atham et al. 2002 catalogue}
\tablenotetext{d}{These quoted distances were found using the formula from Beers et al. 2000.}
\end{deluxetable}

\clearpage

\begin{deluxetable}{lcccccccc}
\tabletypesize{\scriptsize}
\tablecaption{Space motions, angular momentum and energies for Kapteyn stars.  The $\sigma$ values quoted are those derived from a Monte Carlo simulation (see text for details).\label{uvwlze}}
\tablewidth{0pt}
\tablehead{
\colhead{Star} & \colhead{$U$} & \colhead{$\sigma_U$} & \colhead{$V$} & \colhead{$\sigma_V$} 
& \colhead{$W$} & \colhead{$\sigma_W$} & \colhead{$L_z$} & \colhead{$E$} \\ 
\colhead{} & \colhead{km s$^{-1}$} & \colhead{} & \colhead{km s$^{-1}$} & \colhead{} & \colhead{km s$^{-1}$} & \colhead{} & \colhead{kpc km s$^{-1}$} & \colhead{10$^{5}$ km$^{2}$s$^{-2}$}
}
\startdata  
  HD 13979 & -18 & 5& -115 &24& -28 &3& 874 & -1.17 \\
         HD 21022 & -24 &11& -249 &44& 5 &19& -278 & -1.20 \\
         HD 110621 & 19 &30& -255 &20& 58 &5& -280 & -1.23 \\
         HD 111721 & 57 &15& -273 &55& -135 &33& -415 & -1.13 \\
         HD 181007 & -89 &19& -301 &63& -51 &12& -608 & -1.18 \\
         HD 181743 & 42 &14& -335 &70& -56 &11& -913 & -1.16 \\
         HD 186478 & -177 &29& -368 &79& -62 &13& -979 & -1.01 \\
         HD 188031 & 162 &11& -309 &66& 16 &12& -701 & -1.08 \\
         HD 193242 & 45 &13& -172 &28& 47 &7& 363 & -1.23 \\
         HD 208069 & 172 &17& -302 &57& 7 &27& -654 & -1.07 \\
         HD 215601 & 12 &5& -198 &42& 7 &11& 162 & -1.25 \\
         HD 215801 & 14 &6& -204 &46& 123 &12& 118 & -1.17 \\
         CD -30 1121 & 15 &4& -141 &24& -43 &9& 667 & -1.19 \\
         CD -62 1346 & -73 &8& -167 &28& -27 &11& 394 & -1.21 \\
         G 18-54 & -55 &20& -335 &42& 1 &31& -908 & -1.17 \\
         G 24-3 & 2 &26& -252 &24& 85 &5& -254 & -1.21 \\
        \enddata
\end{deluxetable}

\clearpage

\begin{deluxetable}{ll}
\tabletypesize{\scriptsize}
\tablecaption{Constants used in Galactic potential calculations. \label{constants}}
\tablewidth{0pt}
\tablehead{
\colhead{Region} & \colhead{Values}
}
\startdata  
Bulge & M$_b$ = 1.12 x 10$^{10}$M$_{\odot}$ \\
           & a$_b$= 0.0 kpc \\
           & b$_b$ = 0.277 kpc \\ \hline
Disk &  M$_d$ = 8.07 x 10$^{10}$M$_{\odot}$ \\
           & a$_d$= 3.7 kpc \\
           & b$_d$ = 0.20 kpc \\ \hline
Dark Halo & M$_h$ = 5 x 10$^{10}$M$_{\odot}$\\
              & d = 6.0 kpc \\ 
              & $\Phi_{\textrm{offset}}$ = -12.3 x 10$^{4}$ \\
           \enddata
           \end{deluxetable}
          
\clearpage

\begin{deluxetable}{ll}
\tabletypesize{\scriptsize}
\tablecaption{Kinematic values of $\omega$ Centauri. \label{omega}}
\tablewidth{0pt}
\tablehead{
\colhead{$\omega$ Centauri} & \colhead{Parameters} 
}
\startdata  
RA & 201.44$^{\circ}$\\
DEC & -47.00$^{\circ}$\\
 $\mu_{\alpha}$ & -5.08 \textrm{m}$''$ yr$^{-1}$\\
 $\mu_{\delta}$ & -3.57 \textrm{m}$''$ yr$^{-1}$ \\
$V$$_{\textrm{rad}}$ & 232.5 km s$^{-1}$ \\
Distance & 4900 pc \\
$U$ & -61 km s$^{-1}$\\
$V$& -36 km s$^{-1}$\\
$W$ & 6 km s$^{-1}$ \\
$L_z$ & -413 kpc km s$^{-1}$\\
Energy & -1.30 x 10$^{5}$ km$^{2} $s$^{-2}$ \\ 
\enddata
\end{deluxetable}

\clearpage

\begin{deluxetable}{lccccc}
\tabletypesize{\scriptsize}
\tablecaption{Atmospheric parameters adopted for individual stars \label{atmosparam}}
\tablewidth{0pt}
\tablehead{
\colhead{Star} & \colhead{V\tablenotemark{a}} & \colhead{T$_{\textrm{eff}}$} & \colhead{log g} & \colhead{$\chi$} &\colhead{[Fe/H]} \\
\colhead{} &\colhead{} & \colhead{($\pm$ 150K)} & \colhead{$\pm$ 0.20}  & \colhead{$\pm$0.25km s$^{-1}$} & \colhead{}  
}
\startdata  
HD 13979 & 9.16 & 5050 & 2.25 &2.35 &-2.48 \\ 
HD 21022 &  9.19  &   4550       &  1.25        &  1.85  & -2.08     \\
HD 110621 & 9.85 & 5850 & 3.80 & 0.90 &-1.41 \\
HD 111721 & 7.85 & 5000 & 2.60 & 1.52 &-1.34 \\
HD 181007 & 9.24 & 5000 & 2.60 & 1.30 &-1.58 \\
HD 181743 & 9.65 & 6050 & 3.35 &1.00 & -1.47 \\
HD 186478 & 8.85 & 5000 & 1.80 & 1.90 &-1.92 \\
HD 188031 & 10.06 & 6400 & 3.75 & 0.95 & -1.13 \\ 
HD 193242 & 9.10 & 5150 & 2.75 & 1.38 &-1.58 \\
HD 208069 & 9.20 & 5250 & 2.25 & 1.80 &-1.57 \\
HD 215601 & 8.39 & 5110 & 1.80 & 1.75 &-1.10 \\
HD 215801 & 9.99 & 6470 & 3.75 & 0.50 &-1.12 \\
CD -30 1121 & 10.30 & 4960 & 2.45 & 1.00 &-1.63 \\
CD -62 1346 & 9.84 & 5150 & 1.95&1.90& -1.39 \\
G 18-54 & 10.66 &6150  & 4.25 &  1.05 &-1.28  \\
G 24-3 & 10.44 &  6550 & 4.00 & 1.60&-1.29 \\ 
\enddata
\tablenotetext{a}{V magnitudes taken from Eggen, 1996}
\end{deluxetable}

\clearpage
\begin{deluxetable}{cccccccccccccc}
\tabletypesize{\scriptsize}
\tablecaption{Line list used to obtain abundances via spectrum synthesis.\label{lines}}
\tablewidth{0pt}
\tablehead{
\colhead{Species} & \colhead{Wavelength} & \colhead{log gf} & \colhead{$\chi$} \\
\colhead{} & \colhead{\AA} & \colhead{}  & \colhead{eV}  
}
\startdata  
Na I & 5889.95 & 0.00 & 0.117\tablenotemark{a}\\
        & 5895.92 & 0.00 & -0.184\tablenotemark{a} \\
Mg I & 5183.60 & 2.72 & -0.180\\
        & 5528.41 & 4.34 & -0.620\\
        & 5711.09 & 4.34 & -1.833\\
Ca I & 6122.22 & 1.88 & -0.409 \\
        & 6162.17 & 1.90 & 0.100 \\
         & 6439.08 & 2.52 & 0.470\\
        & 6490.84 & 5.80 & -4.745\\
Cu I & 5105.55 & 1.39 & -1.516\\
      & 5782.03 & 1.64 & -3.432\tablenotemark{a}\\
Zr II & 4208.98 & 0.713 & -0.460\\
Ba II & 5853.68 & 0.60 & -1.010\tablenotemark{a}\\
      & 6141.72 & 0.704 & -0.503\tablenotemark{a}\\
      & 6496.90 & 0.604 & -0.380\tablenotemark{a}\\
  \enddata
  \tablenotetext{a} {Hyperfine splitting components used.}
    \end{deluxetable}

  \clearpage
\begin{deluxetable}{cccccccccccccc}
\tabletypesize{\scriptsize}
\tablecaption{Average element abundances of Na, Mg, Ca, Cu, Zr and Ba for all stars in the sample.\label{abundances}}
\tablewidth{0pt}
\tablehead{
\colhead{Star} & \colhead{[Fe/H]} & \colhead{[Na/Fe]} & \colhead{$\sigma$} &
\colhead{[Mg/Fe]} & \colhead{$\sigma$} & \colhead{[Ca/Fe]}  & \colhead{$\sigma$}  & \colhead{[Cu/Fe]}  & \colhead{$\sigma$} & \colhead{[Zr/Fe]}  & \colhead{$\sigma$} & \colhead{[Ba/Fe]}  & \colhead{$\sigma$}
}
\startdata  
HD 21022	&	-2.08	& 0.28 & 0.11& 	0.57	&	0.12	&	0.20	&	0.17	&...		&...		& 0.35 & 0.15& 	0.18	&	0.25	\\
HD 110621	&	-1.40	& 0.38 & 0.10& 	0.58	&	0.10	&	0.46	&	0.24	&...		&...		& 0.25 & 0.15& 	0.30	&	0.14	\\
HD 111721	&	-1.34	& 0.13 & 0.10& 	0.62	&	0.10	&	0.49	&	0.13	&	-0.47	&	0.39	& 0.30 & 0.15& 	0.28	&	0.14	\\
HD 181007	&	-1.58	& 0.35 &0.10 & 	0.45	&	0.10	&	0.26	&	0.24	&	-0.60&	0.14	& 0.35 & 0.15& 	0.52	&	0.10	\\
HD 181743	&	-1.45	& 0.25 &0.10 & 	0.60	&	0.29	&	0.28	&	0.17	&...		&...		& 0.45 & 0.15& 	0.25	&	0.10	\\
HD 186478	&	-1.92	& 0.23 & 0.18& 	0.45	&      0.10 &	0.19	&	0.10	&...		&...		& ... & ...& 	0.23	&	0.10	\\
HD 188031	&	-1.36	& 0.38 &0.10 & 	0.37	&	0.29	&	0.57	&	0.10	&...		&...		& 0.40 & 0.15& 	-0.10	&	0.10	\\
HD 193242	&	-1.58	& -0.13 &0.10 & 	0.23	&	0.31	&	0.12	&	0.25	&...		&...		& 0.25 & 0.15& 	0.14	&	0.10	\\
HD 208069	&	-1.57	& 0.28 & 0.10& 	0.48	&	0.33	&	0.39	&	0.16	&...		&...		& ... &... & 	0.15	&	0.10 \\
HD 215601	&	-1.10	& -0.15 &0.10 & 	0.73	&	0.10	&	0.22	&	0.10	&	-0.60	&0.10		& 0.15 &0.15 & 	-0.03	&	0.25	\\
HD 215801	&	-1.05	& -0.15 &0.10 & 	-0.05	&	0.35	&	...	&	...	&...		&...		& ... & ...& 	...	&...		\\
CD -62.1346	&	-1.53	& 0.13 & 0.10& 	0.32	&	0.10&0.15&0.10&-0.55&0.10& 0.30 &0.15 & 0.95&0.10\\ G 18-54	        &	-1.28	& -0.05 &0.10 & 	0.18	&	0.10	&	0.09	&	0.10	&...		&	...	& 0.05 &0.15 & 	0.19	&	0.10	\\
G 24-3    &	-1.29	& 0.10 &0.10 &	0.50	&	0.10	&	0.08	&	0.18	&...		&	...	& 0.00 &0.15 & 	0.06	&	0.10	\\\hline
HD 13979	&	-2.48	& -0.15 &0.10 & 	-0.03	&	0.39	&	0.10	&	0.14	&	...	&...		& 0.10 & 0.15&	-0.03	&	0.10	\\
CD -30.1121	&	-1.69	& -0.20 & 0.14& 	0.27	&	0.21&0.11&0.10&-0.70&0.10& 0.15 &0.15 & 0.12&0.10\\
\enddata
\end{deluxetable}

\clearpage

\begin{deluxetable}{lccc}
\tabletypesize{\scriptsize}
\tablecaption{Sensitivity of derived log $\epsilon$ abundances to chosen atmospheric parameters\label{abundsens}}
\tablewidth{0pt}
\tablehead{
\colhead{Species} & \colhead{$\Delta$ T$_{\textrm{eff}}$} & \colhead{$\Delta$ log g} & \colhead{$\Delta$ $\chi$} \\
\colhead{} & \colhead{(-200K)} & \colhead{(-0.30)}  & \colhead{(-0.30 km s$^{-1}$)}
}
\startdata
Na I & -0.20 & +0.07 & -0.12 \\
Mg I & -0.13 & +0.03 & -0.06 \\
Ca I & -0.13 & +0.03 & -0.17 \\
Cu I & -0.19 & 0.00 & 0.00 \\
Zr II & -0.09 & -0.09 & -0.17 \\
Ba II & -0.15 & -0.08 & -0.28 \\
\enddata
\end{deluxetable}

\clearpage

\begin{deluxetable}{cccc}
\tabletypesize{\scriptsize}
\tablecaption{Mean values, standard deviations and standard deviations of the means for all abundance arrays. \label{meansstds}}
\tablewidth{0pt}
\tablehead{
\colhead{Species} &  \colhead{Kap} & \colhead{Omega} & \colhead{Halo}\\
\colhead{} & \colhead{(N=14)} & \colhead{(N=50)} & \colhead{(N=275)}
}
\startdata
$<[$Na/Fe$]>$  &     0.15  &0.22 &    -0.07\\
           $\sigma$              &0.18&    0.38        & 0.23 \\
$\sigma_{mean}$&0.050 & 0.054 & 0.014 \\ \hline
$<[$Mg/Fe$]>$   &   0.43 & 0.40   &  0.33\\
        $\sigma $    & 0.19&0.22 &  0.21\\
$\sigma_{mean}$ &0.050 & 0.056&0.013 \\\hline
$<[$Ca/Fe$]>$   &  0.26 & 0.38 &   0.28\\
        $\sigma $    &0.15 &0.11 & 0.16 \\
$\sigma_{mean}$ & 0.041&0.015 &0.010 \\\hline
$<[$ls/Fe$]>$      & 0.24 & 0.20  &   -0.17\\
        $\sigma$     &0.14 & 0.27& 0.27 \\
$\sigma_{mean}$ & 0.038&0.038 & 0.016\\\hline
$<[$hs/Fe$]>$  &   0.21 & 0.25  &   -0.09\\
        $\sigma $    &0.26 &0.41 & 0.48 \\
$\sigma_{mean}$ & 0.070&0.058& 0.029\\
\enddata
\end{deluxetable}

\clearpage

\begin{deluxetable}{lrccc}
\tabletypesize{\scriptsize}
\tablecaption{The results of a parametric t-test on the means and standard deviations of the different populations from the Kapteyn group, the halo stars and previous Omega Cen studies.  If the p-value is $``$small" (i.e. p$<$ 0.05) then the two groups are considered to be from different populations.  If p is larger than 0.05, the test is inconclusive.  Those values shown in bold are the stellar groups that statistically appear to come from different populations.  \label{ttestresults}}
\tablewidth{0pt}
\tablehead{
\colhead{Species} & \colhead{} & \colhead{Kap : Halo} & \colhead{Kap : Omega} & \colhead{Omega : Halo}
}
\startdata
$[$Na/Fe$]$ &t &   3.078  &0.985 &    3.796\\
      & p &\textbf{0.003}&0.332 &\textbf{0.000} \\ \hline
$[$Mg/Fe$]$& t   &   1.965 & 0.516   &  2.063\\
      & p & 0.069& 0.612& \textbf{0.044}\\\hline
$[$Ca/Fe$]$& t   &  0.483 & 2.791 &   5.366\\
      & p &0.645&\textbf{0.013}&\textbf{0.000} \\\hline
$[$ls/Fe$]$    &t  & 9.096 & 0.689  &   8.009\\
      & p &\textbf{0.000} &0.495 & \textbf{0.000}\\\hline
$[$hs/Fe$]$& t  &   3.933 & 0.442  &   5.155\\
      & p & \textbf{0.000}& 0.662& \textbf{0.000}\\\hline
\enddata
\end{deluxetable}

\clearpage

\begin{deluxetable}{lrccc}
\tabletypesize{\scriptsize}
\tablecaption{The results of a non-parametric Kolmogorov-Smirnov test on the different populations from the Kapteyn group, the halo stars and previous Omega Cen studies.  If P is $``$small" (i.e. P $<$ 0.05) then the two groups are considered to be from different populations.  If P is larger than 0.05, the test is inconclusive.  As for Table \ref{ttestresults}, those values shown in bold are the stellar groups that statistically appear to come from different populations.  
\label{KSresults}}
\tablewidth{0pt}
\tablehead{
\colhead{Species} & \colhead{} & \colhead{Kap : Halo} & \colhead{Kap : Omega} & \colhead{Omega : Halo}
}
\startdata
$[$Na/Fe$]$ &D &     0.5304  &0.3200 &    0.4212\\
      & P & \textbf{0.000}  & 0.152     & \textbf{0.000}\\\hline
$[$Mg/Fe$]$& D   &   0.3220 & 0.1933   &  0.2255\\
       & P &0.084 &  0.734  &    \textbf{0.020}\\\hline
$[$Ca/Fe$]$& D   &  0.2504 & 0.5543 &   0.3585\\
         &P&0.329 &  \textbf{0.001} &    \textbf{0.000}\\\hline
$[$ls/Fe$]$    &D  & 0.7814 & 0.1905  &   0.6307\\
       &P & \textbf{0.000} &  0.849 &     \textbf{0.000}\\\hline
$[$hs/Fe$]$& D  &   0.4166 & 0.3771  &   0.4299\\
      & P & \textbf{0.010} &  0.066   &   \textbf{0.000}\\\hline
      \enddata
\end{deluxetable}

\end{document}